\documentclass[aps,showpacs]{revtex4}
\usepackage{array}
\usepackage{float}
\usepackage{amsmath}

\providecommand{\tabularnewline}{\\}

\usepackage{esint}

\begin{document}

\preprint{This line only printed with preprint option}

\title{Canonically quantized soliton in the bound state approach\\
 to heavy baryons in the Skyrme model}

\author{V. Regelskis}

\email{vidas@itpa.lt}

\author{E. Norvai\v{s}as}

\email{norvaisas@itpa.lt}

\affiliation{Institute of Theoretical Physics and Astronomy of Vilnius University,\\
 Go\v{s}tauto 12, Vilnius 01108, Lithuania}

\begin{abstract}
The bound state extension of Skyrme's topological soliton model for
the heavy baryons is quantized canonically in arbitrary reducible
representations of the SU(3) flavor group. The canonical quantization
leads to an additional negative mass term, which stabilizes the quantized
soliton solution. The heavy flavor meson in the field of the soliton
is treated with semiclassical quantization. The representation dependence
of the calculated spectra for the strange, charm and bottom baryons
is explored and compared to the extant empirical spectra. 
\end{abstract}


\maketitle

\section{introduction}

Skyrme's topological soliton model for the baryons \cite{skyrme1}
is a chiral symmetric mesonic representation of QCD in the limit of
large $N_{C}$. In this limit baryons are constructed as topological
soliton solutions to the effective chiral meson Lagrangian. The baryon
number is the second Chern number of the meson field \cite{manton1}).
In the semiclassical version of the model the mass of the pion, which
is the Goldstone boson of the spontaneously broken chiral symmetry,
is introduced through adding a complementary symmetry breaking term
to the Skyrme Lagrangian.

The model yields a qualitative description of the lower energy part
of the spectra of the nucleons and delta resonances, but its direct
extension to describe strange, or heavier, baryons fails phenomenologically
because of the badly broken SU(3) symmetry \cite{manohar1}. In the
chiral limit all quarks are massless, which implies a model with unbroken
SU(3) and higher flavor symmetry groups. The reason is that the flavor
quantum numbers are associated with rotations of collective coordinates,
although the collective coordinates approach is associated with unbroken
flavor symmetry. When the symmetry is badly broken, the rigid rotator
treatment fails, with poor phenomenology as a consequence \cite{prasz}.
A phenomenologically more successful alternative approach of treating
hyperons as bound states of Skyrme solitons and non-topological fields
of K mesons was introduced by Callan and Klebanov \cite{klebanov1,klebanov2}.
In this approach the hyperons with strange and heavier flavor number
appear as bound states of mesons with the appropriate quantum number
and the soliton. In the harmonic approximation the meson-soliton system
is described by a linear wave equation.

When the meson field is expanded perturbatively the model Lagrangian
separates into two parts corresponding soliton and meson field, which
interacts with the soliton field. We show that canonical quantization
of the soliton, which respects the non-commutativity of quantum variables
leads to quantum stabilizing term, which lowers the soliton mass.
The dependence of the meson-soliton interaction on the representation
of the heavy flavor SU(3) is analyzed here in detail, with emphasis
on the Wess-Zumino and the symmetry breaking terms. It is found that
new bound states with higher excitation number can appear in higher
representations of SU(3) group, which are absent in the fundamental
representation. In general the ordering of the calculated spectra
of the heavy flavor baryons agree with the extant empirical values.

The model Lagrangian and the traditional treatment of the kaon fields
that represent heavier meson fields in bound state approach is reviewed
in Sec. II. The generalization of the bound state Skyrme model to
representations of arbitrary dimension is also presented in this section.
In Sec III. the soliton is quantized canonically ab initio in the
collective coordinates framework and embedded into bound-state model
Lagrangian. Sec. IV. contains the wave equation for the meson field
in arbitrary representations and the diagonalization of the Hamiltonian
density in terms of creation-annihilation operators to derive final
model Hamiltonian. Sec V. contains the physical interpretation of
the hyperons states that arise in the model along with a comparison
of the the calculated spectra to experimental data. Sec. VI contains
a summarizing discussion. The notation employed for the SU(3) group
algebra is given in the Appendix A.

\section{Model Lagrangian}

\subsection{The bound state model}

The action of Skyrme model extension to the hyperons is \begin{equation}
S=\int d^{4}x\left(\mathcal{L}_{Sk}+\mathcal{L}_{SB}\right)+S_{WZ}.\end{equation}
 Here $\mathcal{L}_{Sk}$ is Lagrangian density of the original Skyrme
model \cite{skyrme1} \begin{eqnarray}
\mathcal{L}_{Sk} & = & -\frac{f_{\pi}^{2}}{4}\mathrm{Tr}\left\{ (\partial_{\mu}\mathbf{U})\mathbf{U}^{\dagger}(\partial^{\mu}\mathbf{U})\mathbf{U}^{\dagger}\right\} \nonumber \\
 &  & +\frac{1}{32e^{2}}\mathrm{Tr}\left\{ \left[(\partial_{\mu}\mathbf{U})\mathbf{U}^{\dagger},(\partial_{\nu}\mathbf{U})\mathbf{U}^{\dagger}\right]^{2}\right\} ,\label{L_sk}\end{eqnarray}
 and $\mathcal{L}_{SB}$ and $S_{WZ}$ are the actions for the chiral
symmetry breaking and the Wess-Zumino terms respectively.

In the limit of unbroken the $SU\left(3\right)$ symmetry the meson
field $\mathbf{U}$ takes values in $SU\left(3\right)$ group algebra.
When $SU\left(3\right)$ symmetry is badly broken due to a large meson
mass term, the physically relevant field configurations will be small
fluctuations into heavy flavor directions around the pionic soliton.

Here we employ the chiral symmetric field ansatz \cite{klebanov1}
\begin{equation}
\mathbf{U}=\sqrt{\mathbf{U}_{\pi}}\mathbf{U}_{K}\sqrt{\mathbf{U}_{\pi}}\label{Ansatz}\end{equation}
 to construct bound states, where \begin{eqnarray}
\sqrt{\mathbf{U}_{\pi}} & = & exp\left(i\frac{1}{f_{\pi}}\left(\hat{J}_{\left(0,1,m\right)}^{\left(1,1\right)}\hat{x}^{\left(m\right)}\right)F\left(r\right)\right),\\
\mathbf{U}_{K} & = & exp\left(i\frac{1}{f_{K}}\hat{J}_{\left(z,1/2,m\right)}^{\left(1,1\right)}\hat{K}_{\left(z,1/2,m\right)}\right).\label{Uk}\end{eqnarray}
 Here the parameters $\left(z,j,m\right)$ are used for the basis
state notation for the canonical chain $SU\left(2\right)\subset SU\left(3\right)$ \cite{klebanov1}.
To a first approximation it suffices to expand the fields only up
to the second order. The contribution of higher order terms, which
describe self-interactions in the meson field have been found to be
small \cite{bj}.

The kaon (the generalization to the D and B mesons is given in ref. \cite{sco1,sco2})
field has the conventional isodoublet structure: \begin{equation}
K^{\dagger}=\begin{pmatrix}K^{-} & \tilde{K}^{0}\end{pmatrix},\qquad K=\binom{K^{+}}{K^{0}}.\end{equation}
 The meson state is a field that takes values in the adjoint representation
of the $SU(3)$ algebra: \begin{equation}
\begin{array}{cc}
K^{-}=\hat{K}_{\left(\frac{1}{2},\frac{1}{2},-\frac{1}{2}\right)},\quad & K^{+}=\hat{K}_{\left(-\frac{1}{2},\frac{1}{2},\frac{1}{2}\right)},\\
\tilde{K}^{0}=-\hat{K}_{\left(\frac{1}{2},\frac{1}{2},\frac{1}{2}\right)},\quad & K^{0}=\hat{K}_{\left(-\frac{1}{2},\frac{1}{2},-\frac{1}{2}\right)}.\end{array}\end{equation}
 The invariance of the soliton under combined spatial and isospin
rotations implies the following condition for the eigenmode expansion
of the meson field: \begin{equation}
K\left(\mathbf{r},t\right)=k\left(r,t\right)Y_{u\, l\, u_{0}}\left(\vartheta,\varphi\right).\end{equation}
 Here $l$ and $u$ represents orbital and total angular momentum
of the meson field respectively.

The radial part $k\left(r,t\right)$ of the meson field can be expressed
in terms of energy eigenvalues as follows: \begin{equation}
k\left(r,t\right)=\sum_{n>0}\left(k_{-n}\left(r\right)e^{i\omega_{-n}t}b_{n}^{\dagger}+k_{n}\left(r\right)e^{-i\omega_{n}t}a_{n}\right).\label{K_expansion}\end{equation}
 Here $a_{n}$ is the annihilation operator for state with strangeness
$S=-1$ and $b_{n}^{\dagger}$ is the creation operator for state
with strangeness $S=1$.

\subsection{Generalization to arbitrary representations}

The pionic field may be expressed in any irreducible $SU\left(3\right)$
group representation as direct sum of irreducible $SU\left(2\right)$
group representations \cite{jurciukonis1}: \begin{equation}
\sqrt{\mathbf{U}_{\pi}}=\sum_{z,j}^{\left(\lambda,\mu\right)}\oplus D^{j}\left(\alpha\right).\end{equation}
 Using the Maurer-Cartan form notation the explicit form of right
chiral current of pionic field can be expressed as: \begin{eqnarray}
\left(\nabla_{k}\sqrt{\mathbf{U}_{\pi}}\right)\sqrt{\mathbf{U}_{\pi}}^{\dagger} & = & \left(i\left(F^{\prime}-\frac{1}{r}\sin F\right)\left(-1\right)^{m}\hat{x}_{-m}\hat{x}_{k}+i\frac{1}{r}\sin F\,\delta_{m,k}\right.\nonumber \\
 &  & \qquad\left.-\frac{2\sqrt{2}}{r}\sin^{2}\frac{F}{2}\begin{bmatrix}1 & 1 & 1\\
-m & k & n\end{bmatrix}\left(-1\right)^{m}\hat{x}_{n}\right)\left\langle \left\vert \hat{J}_{\left(0,1,m\right)}^{\left(1,1\right)}\right\vert \right\rangle .\label{Maurer-Cartan form}\end{eqnarray}

After substitution of the bound state ansatz (\ref{Ansatz}) into
(\ref{L_sk}) and using (\ref{Maurer-Cartan form}) for pion field
and (\ref{Uk}) for the meson field the Skyrme Lagrangian for the
meson field in a $u=\frac{1}{2}$ state takes the form below after
division by the factor \begin{equation}
N=\frac{1}{4}\dim\left(\lambda,\mu\right)C_{2}\left(\lambda,\mu\right).\label{Norm_factor}\end{equation}
 \begin{eqnarray}
L_{Sk} & = & -M_{cl}+\frac{1}{4f_{K}^{2}}\frac{f_{\pi}}{e}\int d\tilde{r}\tilde{r}^{2}\left(f\,\dot{\tilde{k}}^{2}-h\,\tilde{k}^{\prime2}+V_{eff}\,\tilde{k}^{2}\right).\end{eqnarray}
 Here $M_{cl}$ is the classical soliton mass: \begin{equation}
M_{cl}=4\pi\frac{f_{\pi}}{e}\int\tilde{r}^{2}d\tilde{r}\,\frac{1}{2}\left(d+2s+s\left(2d+s\right)\right).\end{equation}
 The coefficients in this expression are: \begin{eqnarray}
h & = & 1+\frac{1}{4}2s,\nonumber \\
f & = & 1+\frac{1}{4}\left(d+2s\right),\nonumber \\
V_{eff} & = & -\frac{1}{4}\left(d+2s\right)-\frac{1}{4}2s\left(2d+s\right)\nonumber \\
 &  & +\frac{1}{\tilde{r}^{2}}\left(1+\frac{1}{4}\left(d+s\right)\right)\left(2c^{2}+\left(1-4c\right)l\left(l+1\right)\right)\nonumber \\
 &  & +\frac{1}{4}\frac{6}{\tilde{r}^{2}}\left(s\left(c^{2}-\left(2c-1\right)l\left(l+1\right)\right)+\frac{d}{d\tilde{r}}\left(\left(c-l\left(l+1\right)\right)\tilde{F}^{\prime}\sin\tilde{F}\right)\right).\end{eqnarray}
 Here the standard notations employed is: $s=\frac{1}{\tilde{r}^{2}}\sin^{2}\tilde{F}$,
$d=\tilde{F}^{\prime2}$, $c=\sin^{2}\frac{\tilde{F}}{2}$. The dimensionless
parameters are: $\tilde{r}=ef_{\pi}r$, $\tilde{F}\equiv F\left(\tilde{r}\right)$
and $\tilde{k}\equiv k\left(\tilde{r},t\right)$.

\subsection{The Wess-Zumino term}

The Wess-Zumino term plays crucial role in bound-state model. It involves
the meson field only with one time derivative. This term splits the
energies of states with strangeness $S=-1$ and $S=1$ from one another.
The Wess-Zumino action is \begin{equation}
S_{WZ}=-\frac{iN_{c}}{240\pi^{2}}\int d^{5}x\,\epsilon^{\mu\nu\alpha\beta\gamma}\text{Tr}\left\{ \mathbf{R}_{\mu}\mathbf{R}_{\nu}\mathbf{R}_{\alpha}\mathbf{R}_{\beta}\mathbf{R}_{\gamma}\right\} .\end{equation}
 For the field ansatz (\ref{Ansatz}) it may be reduced to the remarkably
elegant expression: \begin{eqnarray}
S_{WZ} & = & \frac{iN_{c}}{2\pi^{2}}\int d^{4}x\frac{1}{4!}\text{Tr}\left\{ \left(\overline{\mathbf{p}}+\overline{\mathbf{p'}}\right)^{3}\overline{\mathbf{k}}\right\} .\label{WZ}\end{eqnarray}
 Here we have used differential 1-form notations: \begin{eqnarray}
\overline{\mathbf{p}} & = & \left(\mathbf{d}\sqrt{\mathbf{U}_{\pi}}\right)\sqrt{\mathbf{U}_{\pi}}^{\dagger},\nonumber \\
\overline{\mathbf{p'}} & = & \sqrt{\mathbf{U}_{\pi}}^{\dagger}\left(\mathbf{d}\sqrt{\mathbf{U}_{\pi}}\right),\nonumber \\
\overline{\mathbf{k}} & = & \left(\mathbf{d}\mathbf{U}_{K}\right)\mathbf{U}_{K}^{\dagger}.\end{eqnarray}
 Explicit evaluation of (\ref{WZ}) upon division by the normalization
factor (\ref{Norm_factor}) yields: \begin{eqnarray}
L_{WZ} & = & -i\frac{C_{3}\left(\lambda,\mu\right)}{C_{2}\left(\lambda,\mu\right)}\frac{3N_{c}}{80f_{K}^{2}}\frac{1}{2\pi^{2}}\int F^{\prime}\sin^{2}F\cdot\left(k^{\dagger}\dot{k}-\dot{k}^{\dagger}k\right)dr.\end{eqnarray}
 In the case of the fundamental $SU(3)$ group representation $\left(1,0\right)$
this result agrees with that in ref. \cite{klebanov2} up to the overall
factor $\frac{1}{16}$, which derives from the different notation
of $f_{K}$ and the present choice of $SU(3)$ group generators.

\subsection{The symmetry breaking term}

The $SU\left(3\right)$ chiral symmetry breaking term of Lagrangian
density is defined as \cite{jurciukonis1} \begin{eqnarray}
\mathcal{L}_{SB} & = & \frac{f_{\pi}^{2}}{4}\left(m_{0}^{2}\text{Tr}\left\{ \mathbf{U}+\mathbf{U}^{\dagger}-2\cdot1\right\} -2m_{8}^{2}\text{Tr}\left\{ \hat{J}_{\left(0,0,0\right)}^{\left(1,1\right)}\left(\mathbf{U}+\mathbf{U}^{\dagger}\right)\right\} \right).\label{L_sb}\end{eqnarray}
 Here (with the exception of the case of the self-adjoint representation)
\begin{eqnarray}
m_{0}^{2} & = & \frac{1}{3}\left(m_{\pi}^{2}+2\frac{f_{K}^{2}}{f_{\pi}^{2}}m_{K}^{2}\right),\\
m_{8}^{2} & = & \frac{10}{3\sqrt{3}}\frac{C_{2}\left(\lambda,\mu\right)}{C_{3}\left(\lambda,\mu\right)}\left(m_{\pi}^{2}-\frac{f_{K}^{2}}{f_{\pi}^{2}}m_{K}^{2}\right).\end{eqnarray}

For the self-adjoint representations $\lambda=\mu$ the symmetry breaking
term is proportional only to $m_{0}^{2}$ because the trace of the
second term is equal to zero.

Substitution of the ansatz (\ref{Ansatz}) into (\ref{L_sb}) leads
to the symmetry breaking term for any irreducible $SU\left(3\right)$
group rep. The explicit form of the general symmetry breaking term
is: \begin{eqnarray}
\mathcal{L}_{SB} & = & \frac{f_{\pi}^{2}}{2}\left(m_{0}^{2}Q_{11}^{\lambda,\mu}-\frac{2}{\sqrt{3}}m_{8}^{2}Q_{12}^{\lambda,\mu}\right)-\frac{f_{\pi}^{2}}{4f_{K}^{2}}\left(m_{0}^{2}Q_{21}^{\lambda,\mu}-\frac{2}{\sqrt{3}}m_{8}^{2}Q_{22}^{\lambda,\mu}\right)K^{^{\dagger}}K.\end{eqnarray}
 Here the notation is \begin{eqnarray}
Q_{11}^{\lambda,\mu} & = & \sum_{z,j,m}^{\lambda,\mu}\cos\left(2mF\left(r\right)\right)-\dim\left(\lambda,\mu\right),\nonumber \\
Q_{12}^{\lambda,\mu} & = & \sum_{z,j}^{\lambda,\mu}\left(\left(\left(\lambda-\mu\right)+3z\right)\sum_{m=-j}^{j}\cos\left(2mF\left(r\right)\right)\right),\nonumber \\
Q_{21}^{\lambda,\mu} & = & \sum_{z,j,m}^{\lambda,\mu}\left(A_{z,j,m}^{\lambda,\mu}\cos\left(2mF\left(r\right)\right)\right),\nonumber \\
Q_{22}^{\lambda,\mu} & = & \sum_{z,j,m}^{\lambda,\mu}\left(\left(\left(\lambda-\mu\right)+3z\right)A_{z,j,m}^{\lambda,\mu}\cos\left(2mF\left(r\right)\right)\right),\end{eqnarray}
 and \begin{eqnarray}
A_{z,j,m}^{\lambda,\mu} & = & -\frac{1}{2j\left(j+1\right)}\left[2j^{3}+j^{4}-mz\left(1+z+\lambda\right)\left(z-\mu-1\right)\right.\nonumber \\
 &  & \qquad-j\left(\lambda-3z^{2}-2z\lambda+m\left(3z+\lambda-\mu\right)+\mu+2z\mu+\lambda\mu\right)\nonumber \\
 &  & \qquad\left.-j^{2}\left(\lambda-3z^{2}-2z\lambda-1+m\left(3z+\lambda+\mu\right)+\mu+2z\mu+\lambda\mu\right)\right].\end{eqnarray}
 For the fundamental $SU(3)$ representation $\left(1,0\right)$ the
symmetry breaking term reduces to the simple form: \begin{eqnarray}
L_{SB} & = & -8\pi\int r^{2}f_{\pi}^{2}m_{\pi}^{2}\sin^{2}\frac{F}{2}dr-\frac{1}{4f_{K}^{2}}\int r^{2}\left(f_{K}^{2}m_{K}^{2}-f_{\pi}^{2}m_{\pi}^{2}\sin^{2}\frac{F}{2}\right)k^{\dag}k\, dr.\end{eqnarray}

For higher representations the weight of symmetry breaking grows,
especially the heavy meson mass part. This term is crucial for phenomenologically
realistic spectra of the heavy flavor hyperons.

\section{Canonical quantization of the soliton}

For the quantization of the Skyrme soliton for arbitrary reducible
$SU\left(3\right)$ group representations $\left(\lambda,\mu\right)$
the quantum operators and soliton field may be separated in the usual
way as: \begin{equation}
\mathbf{U}\left(\mathbf{r},\mathbf{q}\left(t\right)\right)=\mathbf{A}\left(\mathbf{q}\left(t\right)\right)\mathbf{U}_{\pi}\left(\mathbf{r}\right)\mathbf{A}^{\dag}\left(\mathbf{q}\left(t\right)\right).\label{AUA}\end{equation}
 The operator $\mathbf{\hat{A}}$ may be expressed as a direct sum
of Wigner $D$ matrices as: \begin{equation}
\mathbf{A}\left(\mathbf{q}\left(t\right)\right)=\sum_{z,j}^{\left(\lambda,\mu\right)}\oplus D^{j}\left(\mathbf{q}\right).\end{equation}
 The three real time dependent parameters $\mathbf{q}\left(t\right)=\left\{ q_{1}\left(t\right),q_{2}\left(t\right),q_{3}\left(t\right)\right\} $
are quantum variables, which represent Euler angles of rotation of
the soliton. Consideration of the Skyrme Lagrangian quantum mechanically
\textit{ab initio}, the generalized coordinates $\mathbf{q}\left(t\right)$
and velocities $\mathbf{\dot{q}}\left(t\right)$ have to satisfy the
commutation relations \cite{fujii1}: \begin{equation}
\left[\dot{q}^{a},q^{b}\right]=-f^{ab}\left(\mathbf{q}\right).\end{equation}
 Here the tensor $f^{ab}\left(\mathbf{q}\right)$ is a function of
generalized coordinates $\mathbf{q}$ only. It is symmetric as a consequence
of the commutation relation $\left[q^{a},q^{b}\right]=0$.

The explicit form of $f^{ab}\left(\mathbf{q}\right)$ can be determined
only after imposition of the quantization conditions. Using Weyl ordering
of the operators, the commutation relation between a generalized velocity
component $\dot{q}^{a}$ and an arbitrary function $G\left(\mathbf{q}\right)$
is given by \begin{equation}
\left[\dot{q}^{a},G\left(\mathbf{q}\right)\right]=-if^{ab}\left(\mathbf{q}\right)\frac{\partial}{\partial q^{b}}G\left(\mathbf{q}\right).\end{equation}

After substituting (\ref{AUA}) into Skyrme soliton Lagrangian (\ref{L_sk})
the dependence on generalized velocities can be expressed as\begin{equation}
L\left(\mathbf{q},\mathbf{\dot{q},}F\right)=\frac{1}{2}a\left(F\right)\dot{q}^{a}g_{ab}\left(\mathbf{q}\right)\dot{q}^{b}+\mathcal{O}\left(\mathbf{\dot{q}}^{0}\right).\end{equation}
 The function $g_{ab}$ is interpreted as a metric tensor which can
be expressed as a product of functions $C_{a}^{\prime(0,1,m)}\left(\mathbf{q}\right)$ \cite{fujii1}:
\begin{eqnarray}
g_{ab}\left(\mathbf{q}\right) & = & -\frac{1}{4}\dim\left(\lambda,\mu\right)C_{2}\left(\lambda,\mu\right)\left(-1\right)^{m}C_{a}^{\prime(0,1,m)}\left(\mathbf{q}\right)C_{b}^{\prime(0,1,-m)}\left(\mathbf{q}\right).\end{eqnarray}
 Here the soliton inertia momentum $a\left(F\right)$ is defined as
\begin{eqnarray}
a\left(F\right) & = & \frac{8\pi}{3e^{3}f_{\pi}}\int\tilde{r}^{2}\sin^{2}\tilde{F}\left(1+\tilde{F}^{\prime2}+\frac{1}{\tilde{r}^{2}}\sin^{2}\tilde{F}\right)d\tilde{r}.\end{eqnarray}
 The canonical momentum $p_{a}$ conjugate to the generalized coordinate
$q^{a}$ is \begin{equation}
p_{a}=\frac{1}{2}a\left(F\right)\left\{ \dot{q}^{b},g_{ba}\left(\mathbf{q}\right)\right\} .\label{p_sol}\end{equation}
 Employment of the canonical commutation relations leads to the explicit
form of the tensor $f^{ab}\left(\mathbf{q}\right)$: \begin{equation}
f^{ab}\left(\mathbf{q}\right)=\left(a\left(F\right)g_{ab}\left(\mathbf{q}\right)\right)^{-1}.\end{equation}
 Following \cite{fujii1} we define the angular momentum operator
\begin{eqnarray}
I_{a} & = & -i\left\{ p^{b},C_{-b}^{\prime(0,1,a)}\left(\mathbf{q}\right)\right\} =\left(-1\right)^{a}\frac{ia\left(F\right)}{2}\left\{ \dot{q}^{b},C_{-b}^{\prime(0,1,a)}\left(\mathbf{q}\right)\right\} ,\label{j_sol}\end{eqnarray}
 which may be recognized as iso-rotation operator of the soliton.

The explicit form of canonically quantized soliton after division
by the factor (\ref{Norm_factor}) becomes: \begin{equation}
\hat{L}=-M_{cl}(F)-\Delta M_{\left(\lambda,\mu\right)}\left(F\right)+\frac{\mathbf{\hat{I}}^{2}}{2a\left(F\right)}+L_{SB\left(\lambda,\mu\right)}.\end{equation}
 Here $\Delta M_{\left(\lambda,\mu\right)}\left(F\right)$ is quantum
mass correction to the classical soliton mass: \begin{widetext} \begin{eqnarray}
\Delta M_{\left(\lambda,\mu\right)}\left(F\right) & = & -\frac{2\pi}{5a^{2}\left(F\right)}\frac{1}{e^{3}f_{\pi}}\nonumber \\
 &  & \quad\times\int d\tilde{r}\,\tilde{r}^{2}\sin^{2}\tilde{F}\cdot\left(5-11\tilde{F}'^{2}-\sin^{2}\tilde{F}\left(16-16\widetilde{F}'^{2}+\frac{3}{2\tilde{r}^{2}}\right)\right.\nonumber \\
 &  & \qquad\left.+3C_{2}\left(\lambda,\mu\right)\left(4\sin\tilde{F}\left(1-\tilde{F}'^{2}\right)+4\tilde{F}'^{2}+\frac{1}{\tilde{r}^{2}}\sin^{2}\tilde{F}\right)\right).\label{deltaM}\end{eqnarray}
 \end{widetext} The corresponding Hamilton operator for the quantum
soliton is finally: \begin{eqnarray}
\hat{H}_{\left(\lambda,\mu\right)}(F) & = & M\left(F\right)+\Delta M_{\left(\lambda,\mu\right)}\left(F\right)+\frac{\hat{\mathbf{I}}^{2}}{2a\left(F\right)}+M_{SB\left(\lambda,\mu\right)}\left(F\right).\end{eqnarray}

The energy of the canonically quantized soliton differs from the semi-classically
quantized soliton by the appearance of the mass correction $\Delta M_{\left(\lambda,\mu\right)}$,
which depends on representation. This mass correction is negative
and lowers energy of quantum soliton. It shows that quantum soliton
is only approximately \char`\"{}rigid body\char`\"{}.

Minimization of the energy of quantum soliton leads to an integro-differential
equation for the quantum chiral angle $F(\tilde{r})$. The boundary
conditions are the same as for classical chiral angle: $F(0)=\pi$,
$F(\infty)=0$. At large distances ($\tilde{r}\rightarrow\infty)$,
quantum chiral angle equation takes the asymptotic form: \begin{equation}
\tilde{r}^{2}\tilde{F}^{\prime\prime}+2\tilde{r}\tilde{F}^{\prime}-(2+\tilde{m}_{eff}^{2}\tilde{r}^{2})\tilde{F}=0,\end{equation}
 here the quantity $\tilde{m}_{eff}^{2}$ is defined as\begin{equation}
\tilde{m}_{eff}^{2}=\tilde{m}_{\pi}^{2}-\frac{e^{4}}{3\tilde{a}\left(F\right)}\left(8\Delta M_{\left(\lambda,\mu\right)}+\frac{2i\left(i+1\right)+3}{\tilde{a}\left(F\right)}\right).\label{m_eff}\end{equation}
 Here we have used the notation $\tilde{m}=\frac{1}{ef_{\pi}}m$ and
$i\left(i+1\right)$ is the eigenvalue of the operator $\hat{\mathbf{I}}^{2}$.
The solution of the asymptotic equation is \begin{equation}
F(\tilde{r})=C\left(\frac{\tilde{m}_{eff}^{2}}{\tilde{r}}+\frac{1}{\tilde{r}^{2}}\right)\exp(-\tilde{m}_{eff}^{2}\tilde{r}),\end{equation}
 where $C$ is an arbitrary constant, which is determined numerically
by tuning the asymptotic solution to the numerical solution.

The requirement for the quantum soliton to be stable with finite mass
sets the restriction $\tilde{m}_{eff}^{2}>0$. The stability of the
quantum soliton is ensured by the term (\ref{deltaM}). Its absence
in the semi-classical approach leads to the instability of the solution \cite{acus1}.
Furthermore, the appearance of the quantum mass part breaks the scale
invariance of the equation of motion, which is the symmetry of the
classical Lagrangian. This shows that Skyrme Lagrangian has an anomaly.
Therefore, the positive parameter $m_{eff}=ef_{\pi}\tilde{m}_{eff}^{2}$
can be interpreted as an effective pion mass.

\subsection{Semi-classical approach to the heavy flavor meson field}

In the semi-classical approach the heavy flavor meson field is described
in the {}``rest frame'' of the background soliton. In this slowly
corotating meson-soliton system the meson effectively becomes an object
with isospin zero and spin effectively equal to the quantum number
$u$  \cite{klebanov1}. Thus a kaon bounded in the $u=\frac{1}{2}$
wave can be effectively treated as a strange quark. The same treatment
is approximately valid for bounded charm and bottom mesons and leads
to bound states describing charm and bottom baryons.

The semi-classical part of the bound state Lagrangian after substituting
(\ref{Ansatz}) into (\ref{L_sk}) for the $u=\frac{1}{2}$ wave meson
and after dividing by the normalization factor (\ref{Norm_factor})
is \begin{eqnarray}
\delta L & = & -\frac{1}{4f_{K}^{2}}\frac{1}{2}\delta a\left(F,k\right)\dot{\alpha}^{2}+\frac{1}{4f_{K}^{2}}\dot{\alpha}_{0}t_{0}\int\tilde{r}^{2}d\tilde{r}\,\left(i\chi\left(\tilde{k}^{\dagger}\dot{\tilde{k}}-\dot{\tilde{k}}^{\dagger}\tilde{k}\right)\right).\end{eqnarray}
 Here \begin{widetext} \begin{eqnarray}
\dot{\alpha}_{m} & = & i\frac{1}{2}\left\{ \dot{q}^{\alpha},C_{\alpha}^{^{\prime}\left(0,1,m\right)}\left(\mathbf{q}\right)\right\} ,\nonumber \\
\delta a\left(F,k\right) & = & \frac{2}{e^{3}f_{\pi}}\int\tilde{r}^{2}d\tilde{r}\,\Biggl(\left(\left(\frac{1}{4}-\frac{1}{3}\sin^{2}\tilde{F}\right)+\frac{1}{16}\left(\tilde{F}^{\prime2}+\frac{2}{\tilde{r}^{2}}\sin^{2}\tilde{F}\right)\right)\tilde{k}^{2}\nonumber \\
 &  & \qquad\qquad\qquad-\sin^{2}F\left(\frac{3}{8}\tilde{F}^{\prime2}\sin^{2}\tilde{F}+\frac{1}{3\tilde{r}^{2}}\sin^{2}\tilde{F}\right)\tilde{k}^{2}\nonumber \\
 &  & \qquad\qquad\qquad-\frac{1}{8}\frac{d}{d\tilde{r}}\left(\tilde{F}'\sin2\tilde{F}\right)\tilde{k}^{2}+\frac{1}{6}\sin^{2}\tilde{F}\left(\text{ }\tilde{k}^{\prime2}+\frac{2l\left(l+1\right)}{\tilde{r}^{2}}\tilde{k}^{2}\right)\Biggr),\nonumber \\
\chi & = & \frac{1}{e^{2}}\Biggl(\frac{1}{4}\frac{\left(-1\right)^{l}}{2l+1}\left(\tilde{F}^{\prime2}\cos\tilde{F}+\frac{2}{\tilde{r}^{2}}\sin^{2}\tilde{F}\right)-\frac{1}{4\tilde{r}^{2}}\left(\frac{3\left(-1\right)^{l}}{2l+1}-1\right)\frac{d}{d\tilde{r}}\left(\tilde{r}^{2}\tilde{F}^{\prime}\sin\tilde{F}\right)\nonumber \\
 &  & \qquad\qquad\qquad+\frac{1}{6}\tilde{F}^{\prime2}\sin^{2}\frac{\tilde{F}}{2}-\frac{l\left(l+1\right)}{3\tilde{r}^{2}}\sin^{2}\tilde{F}+\frac{2}{3}\sin^{2}\frac{\tilde{F}}{2}+\frac{\left(-1\right)^{l}}{2l+1}\cos\tilde{F}\Biggr).\end{eqnarray}
 \end{widetext} Here $\delta a\left(F,k\right)$ is a small positive
parameter, which can be interpreted as an additional contribution
to the soliton inertion momentum. This shows that the background field
rotating together with quantum soliton slows it down somewhat. The
operators multiplied with $\chi\left(\tilde{r}\right)$ are responsible
for the spin-spin interaction of soliton and the bounded meson. These
operators allow discrimination between states with different total
spin.

At the semi-classical level the Wess-Zumino action

\begin{eqnarray}
S_{WZ} & = & -\frac{iN_{c}}{2\pi^{2}}\int_{M}\frac{1}{4!}\text{Tr}\left\{ \mathbf{\overline{w}}^{3}\left(\overline{\alpha}'+W^{\dagger}\mathbf{\overline{\alpha}}'W\right)\right\} ,\end{eqnarray}
 contributes to the spin-spin interaction between the soliton and
the heavy flavor meson. Here $\overline{\mathbf{w}}=\mathbf{d}UU^{\dagger}$
and $\mathbf{\overline{\alpha}}'=\alpha'^{\dagger}\mathbf{d}\alpha'$
are 1-forms. The semi-classical part of Wess-Zumino Lagrangian can
be written in the compact form \begin{eqnarray}
\delta L_{WZ} & = & \dot{\alpha}_{0}t_{0}\frac{1}{4f_{K}^{2}}\int\tilde{r}^{2}d\tilde{r}\,\chi_{WZ}\left(\tilde{r},\lambda,\mu\right)\tilde{k}^{2},\end{eqnarray}
 where\begin{widetext}\begin{eqnarray}
\chi_{WZ}\left(\tilde{r},\lambda,\mu\right) & = & \frac{C_{3}\left(\lambda,\mu\right)}{C_{2}\left(\lambda,\mu\right)}\frac{1}{10\pi^{2}}\frac{\tilde{F}'}{\tilde{r}^{2}}\nonumber \\
 &  & \times\biggl(\sin^{2}\frac{\tilde{F}}{2}\left(3+5\cos\tilde{F}+4\cos2\tilde{F}-2\left(\cos\tilde{F}-2\cos2\tilde{F}\right)\right)\nonumber \\
 &  & \qquad\qquad\qquad-l\left(l+1\right)\left(2\left(\cos\tilde{F}+\cos2\tilde{F}\right)-\cos3\tilde{F}-4\right)\nonumber \\
 &  & \qquad\qquad\qquad+l\left(l+1\right)\frac{1}{2}\left(\cos\tilde{F}+4\cos2\tilde{F}-3\cos3\tilde{F}\right)\biggr).\end{eqnarray}
 \end{widetext}

\subsection{The bound state mechanics}

The final expression of the semi-classical Lagrangian is \begin{widetext}
\begin{eqnarray}
L & = & -M_{cl}-\Delta M-M_{SB}+\frac{1}{2}\left(a\left(F\right)+\delta a\left(F,k\right)\right)\dot{\alpha}^{2}\nonumber \\
 &  & +\frac{f_{\pi}}{e}\int\tilde{r}^{2}d\tilde{r}\,\left(f\,\dot{\tilde{k}}^{2}+i\Lambda\left(\tilde{r},\lambda,\mu\right)\left(\tilde{k}^{\dagger}\dot{\tilde{k}}-\dot{\tilde{k}}^{\dagger}\tilde{k}\right)-h\,\tilde{k}^{\prime2}-\left(M_{K}^{2}\left(\tilde{r},\lambda,\mu\right)+V_{eff}\,\right)\tilde{k}^{2}\right)\nonumber \\
 &  & +\dot{\alpha}_{0}u_{0}\int\tilde{r}^{2}d\tilde{r}\,\left(i\chi\left(\tilde{k}^{\dagger}\dot{\tilde{k}}-\dot{\tilde{k}}^{\dagger}\tilde{k}\right)+\chi_{WZ}\left(\tilde{r},\lambda,\mu\right)\tilde{k}^{2}\right).\label{L_full}\end{eqnarray}
 \end{widetext} Here the coefficients coming from the Wess-Zumino
and symmetry breaking terms depend on the chosen representation: \begin{eqnarray}
\Lambda\left(\tilde{r},\lambda,\mu\right) & = & -\frac{e^{2}}{4}\frac{3}{5}\frac{C_{3}\left(\lambda,\mu\right)}{C_{2}\left(\lambda,\mu\right)}\frac{N_{c}}{2\pi^{2}\tilde{r}^{2}}\tilde{F}^{\prime}\sin^{2}\tilde{F},\\
M_{\pi}^{2}\left(\tilde{r},\lambda,\mu\right) & = & -\frac{1}{4}\left(2\tilde{m}_{0}^{2}Q_{11}^{\lambda,\mu}-\frac{4}{\sqrt{3}}\tilde{m}_{8}^{2}Q_{12}^{\lambda,\mu}\right),\\
M_{K}^{2}\left(\tilde{r},\lambda,\mu\right) & = & \frac{1}{4}2\left(2\tilde{m}_{0}^{2}Q_{21}^{\lambda,\mu}-\frac{4}{\sqrt{3}}\tilde{m}_{8}^{2}Q_{22}^{\lambda,\mu}\right).\end{eqnarray}
 In the case of the fundamental $SU(3)$ group representation these
take the simple forms \begin{eqnarray*}
M_{\pi}^{2}\left(\tilde{r},1,0\right) & = & 2\tilde{m}_{\pi}^{2}\text{ }\sin^{2}\frac{\tilde{F}}{2},\\
M_{K}^{2}\left(\tilde{r},1,0\right) & = & -\tilde{m}_{\pi}^{2}\sin^{2}\frac{\tilde{F}}{2}+\frac{f_{K}^{2}}{f_{\pi}^{2}}\tilde{m}_{K}^{2},\\
\Lambda\left(\tilde{r},1,0\right) & = & \frac{e^{2}}{4}N_{c}\left(-\frac{1}{2\pi^{2}}\frac{1}{\tilde{r}^{2}}\sin^{2}\tilde{F}\cdot\tilde{F}^{\prime}\right).\end{eqnarray*}

The equation of motion for the meson field corresponding Lagrangian
(\ref{L_full}) is \begin{widetext} \begin{eqnarray}
0 & = & -f\,\ddot{\tilde{k}}+2i\left(\Lambda\left(\tilde{r},\lambda,\mu\right)+\frac{e}{f_{\pi}}\dot{\alpha}_{0}u_{0}\chi\right)\dot{\tilde{k}}\nonumber \\
 &  & -\left(V_{eff}+M_{K}^{2}\left(\tilde{r},\lambda,\mu\right)-\frac{e}{f_{\pi}}\dot{\alpha}_{0}u_{0}\chi_{WZ}\left(\tilde{r},\lambda,\mu\right)\right)\tilde{k}+\frac{1}{\tilde{r}^{2}}\left(\tilde{r}^{2}\, h\,\tilde{k}^{\prime}\right)^{\prime}\,.\end{eqnarray}
 \end{widetext} Here a term $\delta a\left(F,k\right)$ has been
dropped, as it is of order $\frac{1}{N_{c}^{2}}$, which is not relevant
here.

After substitution of (\ref{K_expansion}) we get two independent
equations that represent states with strangeness $S=-1$ and $S=1$
respectively emerge: \begin{widetext} \begin{eqnarray}
 &  & \left(\tilde{\omega}_{n}^{2}f+2\tilde{\omega}_{n}\left(\Lambda+\frac{e}{f_{\pi}}\dot{\alpha}_{0}u_{0}\chi\right)-V_{eff}-M_{K}^{2}+\frac{e}{f_{\pi}}\dot{\alpha}_{0}u_{0}\chi_{WZ}\right)\tilde{k}_{n}\nonumber \\
 &  & +\frac{1}{\tilde{r}^{2}}\left(\tilde{r}^{2}\, h\,\tilde{k}_{n}^{\prime}\right)^{\prime}=0,\\
 &  & \left(\tilde{\omega}_{-n}^{2}f-2\tilde{\omega}_{-n}\left(\Lambda+\frac{e}{f_{\pi}}\dot{\alpha}_{0}u_{0}\chi\right)-V_{eff}-M_{K}^{2}+\frac{e}{f_{\pi}}\dot{\alpha}_{0}u_{0}\chi_{WZ}\right)\tilde{k}_{-n}\nonumber \\
 &  & +\frac{1}{\tilde{r}^{2}}\left(\tilde{r}^{2}\, h\,\tilde{k}_{-n}^{\prime}\right)^{\prime}=0.\end{eqnarray}
 \end{widetext} Here $\tilde{\omega}=\frac{1}{ef_{\pi}}\omega$.

Upon diagonalization of the Hamiltonian in form of energy eigenvalues
we find that the canonical momentum conjugate to $\tilde{k}$ is:
\begin{equation}
\tilde{\pi}_{m}=f\,\dot{\tilde{k}}_{m}^{^{\dagger}}+i\left(\Lambda+\frac{e}{f_{\pi}}\dot{\alpha}_{0}u_{0}\chi\right)\tilde{k}_{m}^{^{\dagger}}.\end{equation}
 The canonical commutation relations lead to the following orthogonality
relations:\begin{widetext} \begin{eqnarray}
\frac{1}{4f_{K}^{2}}\int d\tilde{r}\,\tilde{r}^{2}\tilde{k}_{n}\tilde{k}_{m}\left(\left(\tilde{\omega}_{n}+\tilde{\omega}_{m}\right)f+2\left(\Lambda+\frac{e}{f_{\pi}}\dot{\alpha}_{0}u_{0}\chi\right)\right) & = & \delta_{nm},\\
\frac{1}{4f_{K}^{2}}\int d\tilde{r}\,\tilde{r}^{2}\tilde{k}_{-n}\tilde{k}_{-m}\left(\left(\tilde{\omega}_{-n}+\tilde{\omega}_{-m}\right)f-2\left(\Lambda+\frac{e}{f_{\pi}}\dot{\alpha}_{0}u_{0}\chi\right)\right) & = & \delta_{nm}.\end{eqnarray}
 \end{widetext} In terms of creation and annihilation operators which
obey the usual algebra \begin{equation}
\left[a_{n},a_{m}^{^{\dagger}}\right]=\delta_{nm},\qquad\left[b_{n},b_{m}^{^{\dagger}}\right]=\delta_{nm},\end{equation}
 the diagonalized Hamilton operator for kaon fields becomes \begin{equation}
\hat{H}=\tilde{\omega}_{n}a_{n}^{^{\dagger}}a_{n}+\tilde{\omega}_{-n}b_{-n}^{\dagger}b_{-n}.\end{equation}

The momentum canonically conjugate to the quantum degrees of freedom
is \begin{eqnarray}
p'_{\alpha} & = & -a'\left(F,k\right)\frac{1}{2}\left\{ \dot{q}^{\beta},g_{\beta\alpha}\left(\mathbf{q}\right)\right\} +iC_{\alpha}^{\prime\left(0,1,0\right)}\left(\mathbf{q}\right)u_{0}\chi'\left(F,k,\lambda,\mu\right),\end{eqnarray}
 where \begin{eqnarray}
a'\left(F,k\right) & = & a\left(F\right)+\delta a\left(F,k\right);\\
\chi'\left(F,k,\lambda,\mu\right) & = & \int\tilde{r}^{2}d\tilde{r}\,\left(i\chi\left(\tilde{k}^{\dagger}\dot{\tilde{k}}-\dot{\tilde{k}}^{\dagger}\tilde{k}\right)+\chi_{WZ}\left(\tilde{r},\lambda,\mu\right)\tilde{k}^{2}\right).\end{eqnarray}
 This differs from (\ref{p_sol}) because of the influence of the
bound meson field. Now we can write the final expression of the angular
momentum operator of soliton rotating with the bounded field: \begin{eqnarray}
I'_{a} & = & \left(-\right)^{a}a'\left(F,k\right)\dot{\alpha}_{a}+\delta_{a,0}t_{0}\chi'\left(F,k,\lambda,\mu\right).\end{eqnarray}
 It differs from (\ref{j_sol}) by additional part which arose because
of the interaction with bounded field. Using the canonical Legendre
transformation \begin{eqnarray}
H & = & \frac{1}{2}\left\{ \dot{q}^{\alpha},p_{\alpha}\right\} +\int\tilde{r}^{2}d\tilde{r}\,\left(\tilde{\pi}\dot{\tilde{k}}+\dot{\tilde{k}}^{\dagger}\tilde{\pi}^{\dagger}\right)-L\end{eqnarray}
 the following Hamiltonian obtains: \begin{eqnarray}
\hat{H} & = & M_{cl}+\Delta M+M_{SB}+\omega+\frac{1}{2a'\left(F,k\right)}\left(\mathbf{\hat{I'}}^{2}-u_{0}^{2}\chi'^{2}\left(F,k,\lambda,\mu\right)\right).\end{eqnarray}

\section{Interpretation of physical states and numerical results}

Quantum states of soliton are identified as isospin states. Total
angular momentum $j$ of soliton is equal to its isospin $\left(i,j\right)=\left(i,i\right).$
Lowest energy state of bounded field has $l=1$ \cite{klebanov1}.
The spin-spin interaction terms let to distinguish states with different
total spin. Thus we recognize final state with $\left(0,1/2\right)=\left(0,0\right)+\left(0,1/2\right)$
to be $\Lambda\left(P_{01}\right)$, state with $\left(1,1/2\right)=\left(1,1\right)+\left(0,1/2\right)$
to be $\Sigma\left(P_{11}\right)$ and state with $\left(1,3/2\right)=\left(1,1\right)+\left(0,1/2\right)$
to be $\Sigma^{*}\left(P_{13}\right)$.

Bound state model has 5 independent parameters - model parameter $e$,
pion decay constant $f_{\pi}$, pion mass $m_{\pi}$, kaon decay constant
$f_{K}$ and kaon mass $m_{K}$ for calculating properties of hyperons.
For calculations of properties of charmed or bottom baryons correspondingly
we have to use decay constant $f_{D}$ or $f_{B}$ and mesons mass
$m_{D}$ or $m_{B}$. All these parameters are measured experimentally
except the model parameter $e$ which value can be calculated by setting
quantum soliton with isospin $i=\frac{1}{2}$ mass equal to the experimentally
measured mass of the nucleon\begin{eqnarray}
M_{N} & = & \frac{f_{\pi}}{e}M\left(\tilde{F}\right)+e^{3}f_{\pi}\left(\Delta M_{\left(\lambda,\mu\right)}\left(\tilde{F}\right)+\frac{3}{8a\left(\tilde{F}\right)}\right)+f_{\pi}^{2}M_{SB\left(\lambda,\mu\right)}\left(\tilde{F}\right).\label{ef_1}\end{eqnarray}
 We have calculated hyperon mass spectra for wave $l=1$ using three
different values of pion mass parameter $m_{\pi}$ in the symmetry
braking term (\ref{L_sb}). We have chosen input parameters to be
$M_{N}=939MeV$, $f_{\pi}=65.35MeV$, $f_{K}/f_{\pi}=1.22$, $m_{K}=495\, MeV$
and three different choices of the pion mass parameter in the symmetry
breaking term: $m_{\pi}=0\, MeV$, $m_{\pi}=71.6\, MeV$ and $m_{\pi}=137\, MeV$.
The results for irreps $\left(1,0\right)$, $\left(2,0\right)$ and
$\left(2,1\right)$ are presented in the table I (the slash means
that we have found no stable solutions). The value $m_{\pi}=0$ means
that pions at classical level are treated as pure Goldstone bosons
but they acquire mass at quantum level. The value $m_{\pi}=137MeV$
represents the classical mass of the pion in the symmetry breaking
term. Although, by setting $m_{\pi}=0$ from (\ref{m_eff}) we get
$m_{eff}=108MeV$ and by setting $m_{\pi}=137$ we get $m_{eff}=193MeV$.
By fitting calculations to the nucleon properties we have found the
value of the symmetry breaking parameter to be $m_{\pi}=71.6MeV$
which leads to the correct effective mass of the pion $m_{eff}=137MeV$.
Consequent calculations results for charmed and bottom baryons for
irrep $\left(1,0\right)$ are given in tables II and III. The input
parameters for charmed baryons are $f_{D}/f_{\pi}=1.7$, $m_{K}=1867\, MeV$
and for bottom baryons are $f_{B}/f_{\pi}=2$, $m_{K}=5279\, MeV$.

It was shown in ref. \cite{acus1} that there's an alternative way
to do the calculations. One have to choose isoscalar part of nucleon
electric mean square radius \begin{equation}
\left\langle r_{E,I=0}^{2}\right\rangle =-\frac{2}{\pi}\frac{1}{\left(ef_{\pi}\right)^{2}}\int\tilde{r}^{2}d\tilde{r}\,\tilde{F}'\sin^{2}\tilde{F}\label{ef_2}\end{equation}
 as input parameter instead of pion decay constant. Now $f_{\pi}$
and $e$ are model parameters and can be calculated from equations
(\ref{ef_1}) and (\ref{ef_2}), although, the ratio $f_{K}/f_{\pi}$
($f_{D}/f_{\pi}$or $f_{B}/f_{\pi}$) is kept fixed as previously.
We choose input parameter to be $\left\langle r_{E,I=0}^{2}\right\rangle =0.604fm^{2}$.
By fitting calculations to the nucleon properties we find the value
of the symmetry breaking parameter to be $m_{\pi}=116MeV$. Calculated
values of parameter $f_{\pi}$ for some irreps and $m_{\pi}$ values
is given in table IV. Hyperons calculations results for wave $l=1$
are given in table V and results for wave $l=0$ are given in table
VI. Calculations of charmed and bottom baryons are given in table
VII and table VIII.

\section{Discussion}

In this paper, we have discussed the bound state model describing
heavy baryons containing a single heavy quark. We have constructed
a bound state out of canonically quantized soliton and heavy meson.
Soliton was quantized canonically in the framework of the collective
coordinates formalism for arbitrary irreducible $SU\left(3\right)$
representation. We have treated soliton field quantum mechanically
\textit{ab initio}. The canonical quantization of the soliton respecting
noncommutativity of quantum variables - collective coordinates which
are the Euler angles of the soliton rotation, leaded to quantum soliton
stabilizing term. This term depends on the representation $\left(\lambda,\mu\right)$
and lowers soliton mass. Bounded meson field was treated semiclassically.
The symmetry breaking and Wess-Zumino terms play a crucial role for
the bounded field and also depend on the representation. For self
adjoint representation $\lambda=\mu$ Wess-Zumino term vanishes and
symmetry breaking term is restricted to $SU\left(2\right)$ symmetry
breaking term. The bound state approach was done precisely respecting
canonical Lagrangian and Hamiltonian formalism.

We found semiclassical Hamiltonian describing bounds states in the
background of the quantum soliton. The representation $\left(\lambda,\mu\right)$
influences the explicit expression of Hamiltonian and tunes effective
Yukawa potential. Consequently, the dependence on representation can
be interpreted as a new discrete phenomenological parameter of the
model. However the explicit physical meaning of the dependence on
representation is not completely understood.

The calculations were done for the spectra of the strange, charm and
bottom baryons, where they were treated as a bound states of quantum
soliton and appropriate flavor meson. The predicted mass values for
the non-excited hyperons are very close to the experimental ones.
Although, the canonical approach in not very successful in describing
excited states. The same remarks are valid for charm and bottom flavored
baryons. However, we were able to investigate charmed and bottom baryons
only in the fundamental $SU\left(3\right)$ rep. Also we put a lower
bound on the ratio $f_{B}/f_{\pi}\geq2$. The energies of calculations
for higher reps are far too high because of rapidly growing influence
of symmetry breaking term. Nevertheless, the results could drastically
change if different mass term would be employed. The mass term is
very important because the heavy meson is treated semiclassicaly.
Therefore, right mass term could lead to complete set of states of
charmed and bottom baryons.

\begin{acknowledgments}
The authors are grateful to prof. Dan-Olof Riska for collaboration
and useful discussions.

\vspace{1cm}

\end{acknowledgments}
\begin{center}
\begin{table}[H]

\begin{centering}
\begin{tabular}{|>{\centering}m{1.7cm}|>{\centering}m{1cm}|>{\centering}m{1cm}|>{\centering}m{1cm}|>{\centering}m{1cm}|>{\centering}m{1cm}|>{\centering}m{1cm}|>{\centering}m{1cm}|>{\centering}m{1cm}|>{\centering}m{1cm}|}
\hline 
 & \multicolumn{3}{c|}{$\Lambda\left(1116\right)$} & \multicolumn{3}{c|}{$\Sigma\left(1193\right)$} & \multicolumn{3}{c|}{$\Sigma^{*}\left(1385\right)$}\tabularnewline
\hline 
irrep.{\LARGE \textbackslash{}}$m_{\pi}$  & 0  & 71.6  & 137  & 0  & 71.6  & 137  & 0  & 71.6  & 137\tabularnewline
\hline 
$\left(1,0\right)$  & 1049  & 1029  & 990  & -  & 1235  & 1201  & -  & 1355  & 1374\tabularnewline
\hline 
$\left(2,0\right)$  & 1015  & 1067  & 1036  & 1221  & 1205  & 1178  & 1330  & 1320  & 1305\tabularnewline
\hline 
$\left(2,1\right)$  & 1082  & 1196  & 1159  & 1330  & 1310  & 1272  & 1425  & 1411  & 1383\tabularnewline
\hline
\end{tabular}
\par\end{centering}

\caption{Hyperon mass spectra with $l=1$, $MeV$.}

\end{table}

\par\end{center}

\begin{center}
\begin{table}[H]

\begin{centering}
\begin{tabular}{|>{\centering}m{1.2cm}|>{\centering}m{1cm}|>{\centering}m{1cm}|>{\centering}m{1cm}|>{\centering}m{1cm}|>{\centering}m{1cm}|>{\centering}m{1cm}|>{\centering}m{1cm}|>{\centering}m{1cm}|>{\centering}m{1cm}|}
\hline 
 & \multicolumn{3}{c|}{$\Lambda_{c}\left(2287\right)$} & \multicolumn{3}{c|}{$\Sigma_{c}\left(2455\right)$} & \multicolumn{3}{c|}{$\Sigma_{c}^{*}\left(2520\right)$}\tabularnewline
\hline 
$m_{\pi}$  & 0  & 71.6  & 137  & 0  & 71.6  & 137  & 0  & 71.6  & 137\tabularnewline
\hline 
 & 2198  & 2155  & 2068  & -  & 2492  & 2355  & -  & 2658  & 2606\tabularnewline
\hline
\end{tabular}
\par\end{centering}

\caption{Charmed baryons spectra for irrep. $\left(1,0\right)$ with $l=1$,
$MeV$.}

\end{table}

\par\end{center}

\begin{center}
\begin{table}[H]

\begin{centering}
\begin{tabular}{|>{\centering}m{1.2cm}|>{\centering}m{1cm}|>{\centering}m{1cm}|>{\centering}m{1cm}|>{\centering}m{1cm}|>{\centering}m{1cm}|>{\centering}m{1cm}|>{\centering}m{1cm}|>{\centering}m{1cm}|>{\centering}m{1cm}|}
\hline 
 & \multicolumn{3}{c|}{$\Lambda_{b}\left(5620\right)$} & \multicolumn{3}{c|}{$\Sigma_{b}\left(5810\right)$} & \multicolumn{3}{c|}{$\Sigma_{b}^{*}\left(5830\right)$}\tabularnewline
\hline 
$m_{\pi}$  & 0  & 71.6  & 137  & 0  & 71.6  & 137  & 0  & 71.6  & 137\tabularnewline
\hline 
 & 5584  & 5485  & 5284  & -  & 6236  & 5865  & -  & 6325  & 6177\tabularnewline
\hline
\end{tabular}
\par\end{centering}

\caption{Bottom baryons spectra for irrep. $\left(1,0\right)$ with $l=1$,
$MeV$.}

\end{table}

\par\end{center}

\begin{center}
\begin{table}[H]

\begin{centering}
\begin{tabular}{|>{\centering}m{1.2cm}|>{\centering}m{1cm}|>{\centering}m{1cm}|>{\centering}m{1cm}|}
\hline 
$m_{\pi}$  & 0  & 116  & 137\tabularnewline
\hline 
$\left(1,0\right)$  & 58  & 54,4  & 53\tabularnewline
\hline 
$\left(2,0\right)$  & 60  & 57,6  & 56,8\tabularnewline
\hline 
$\left(2,1\right)$  & 61,8  & 60  & 58,5\tabularnewline
\hline
\end{tabular}
\par\end{centering}

\caption{Calculated $f_{\pi}$, $MeV$.}

\end{table}

\par\end{center}

\begin{center}
\begin{table}[H]

\begin{centering}
\begin{tabular}{|>{\centering}m{1.7cm}|>{\centering}m{1cm}|>{\centering}m{1cm}|>{\centering}m{1cm}|>{\centering}m{1cm}|>{\centering}m{1cm}|>{\centering}m{1cm}|>{\centering}m{1cm}|>{\centering}m{1cm}|>{\centering}m{1cm}|}
\hline 
 & \multicolumn{3}{c|}{$\Lambda\left(1116\right)$} & \multicolumn{3}{c|}{$\Sigma\left(1193\right)$} & \multicolumn{3}{c|}{$\Sigma^{*}\left(1385\right)$}\tabularnewline
\hline 
irrep.{\LARGE \textbackslash{}}$m_{\pi}$  & 0  & 116  & 137  & 0  & 116  & 137  & 0  & 116  & 137\tabularnewline
\hline 
$\left(1,0\right)$  & 1128  & 1098  & 1086  & -  & 1202  & 1190  & -  & 1324  & 1318\tabularnewline
\hline 
$\left(2,0\right)$  & 1118  & 1089  & 1078  & 1231  & 1191  & 1180  & 1319  & 1287  & 1278\tabularnewline
\hline 
$\left(2,1\right)$  & 1223  & 1186  & 1174  & 1327  & 1280  & 1266  & 1414  & 1375  & 1364\tabularnewline
\hline
\end{tabular}
\par\end{centering}

\caption{Hyperon mass spectra with $l=1$, $MeV$.}

\end{table}

\par\end{center}

\begin{center}
\begin{table}[H]

\begin{centering}
\begin{tabular}{|>{\centering}m{1.7cm}|>{\centering}m{1cm}|>{\centering}m{1cm}|>{\centering}m{1cm}|>{\centering}m{1cm}|>{\centering}m{1cm}|>{\centering}m{1cm}|>{\centering}m{1cm}|>{\centering}m{1cm}|>{\centering}m{1cm}|}
\hline 
 & \multicolumn{3}{c|}{$\Lambda\left(1405\right)$} & \multicolumn{3}{c|}{$\Sigma\left(1660\right)$} & \multicolumn{3}{c|}{$\Sigma^{*}\left(1670\right)$}\tabularnewline
\hline 
irrep.{\LARGE \textbackslash{}}$m_{\pi}$  & 0  & 116  & 137  & 0  & 116  & 137  & 0  & 116  & 137\tabularnewline
\hline 
$\left(1,0\right)$  & -  & 1288  & 1278  & -  & 1479  & -  & -  & 1418  & -\tabularnewline
\hline 
$\left(2,0\right)$  & 1249  & -  & -  & -  & -  & -  & -  & -  & -\tabularnewline
\hline 
$\left(2,1\right)$  & 1307  & -  & -  & -  & 1446  & -  & -  & 1355  & -\tabularnewline
\hline
\end{tabular}
\par\end{centering}

\caption{Hyperon mass spectra with $l=0$, $MeV$.}

\end{table}

\par\end{center}

\begin{center}
\begin{table}[H]

\begin{centering}
\begin{tabular}{|>{\centering}m{1.2cm}|>{\centering}m{1cm}|>{\centering}m{1cm}|>{\centering}m{1cm}|>{\centering}m{1cm}|>{\centering}m{1cm}|>{\centering}m{1cm}|>{\centering}m{1cm}|>{\centering}m{1cm}|>{\centering}m{1cm}|}
\hline 
 & \multicolumn{3}{c|}{$\Lambda_{c}\left(2287\right)$} & \multicolumn{3}{c|}{$\Sigma_{c}\left(2455\right)$} & \multicolumn{3}{c|}{$\Sigma_{c}^{*}\left(2520\right)$}\tabularnewline
\hline 
$m_{\pi}$  & 0  & 116  & 137  & 0  & 116  & 137  & 0  & 116  & 137\tabularnewline
\hline 
$l=1$  & 2343  & 2251  & 2219  & -  & 2374  & 2328  & -  & 2563  & 2531\tabularnewline
\hline 
 & \multicolumn{3}{c|}{$\Lambda_{c}\left(2593\right)$} & \multicolumn{3}{c|}{$\Sigma_{c}\left(2800\right)$} & \multicolumn{3}{c|}{}\tabularnewline
\hline 
$l=0$  & -  & 2468  & 2435  & -  & -  & -  &  &  & \tabularnewline
\hline
\end{tabular}
\par\end{centering}

\caption{Charmed baryons spectra for rep. $\left(1,0\right)$, $MeV$.}

\end{table}

\par\end{center}

\begin{center}
\begin{table}[H]

\begin{centering}
\begin{tabular}{|>{\centering}m{1.2cm}|>{\centering}m{1cm}|>{\centering}m{1cm}|>{\centering}m{1cm}|>{\centering}m{1cm}|>{\centering}m{1cm}|>{\centering}m{1cm}|>{\centering}m{1cm}|>{\centering}m{1cm}|>{\centering}m{1cm}|}
\hline 
 & \multicolumn{3}{c|}{$\Lambda_{b}\left(5620\right)$} & \multicolumn{3}{c|}{$\Sigma_{b}\left(5810\right)$} & \multicolumn{3}{c|}{$\Sigma_{b}^{*}\left(5830\right)$}\tabularnewline
\hline 
$m_{\pi}$  & 0  & 116  & 137  & 0  & 116  & 137  & 0  & 116  & 137\tabularnewline
\hline 
$l=1$  & 5824  & 5533  & 5436  & -  & 5797  & 5659  & -  & 6020  & 5900\tabularnewline
\hline 
 & \multicolumn{3}{c|}{$\Lambda_{b}\left(?\right)$} & \multicolumn{3}{c|}{} & \multicolumn{3}{c|}{}\tabularnewline
\hline 
$l=0$  & -  & 5770  & 5673  &  &  &  &  &  & \tabularnewline
\hline
\end{tabular}
\par\end{centering}

\caption{Bottom baryons spectra for rep. $\left(1,0\right)$, $MeV$.}

\end{table}

\par\end{center}

\appendix

\section{Elements of $SU\left(3\right)$ group algebra}

The $SU\left(3\right)$ group generators are defined as components
of irreducible $\left(1,1\right)$ tensors. Their relation to the
Gell-Mann generators $\Lambda_{k}$ are: \begin{equation}
\begin{array}{llllll}
J_{(0,0,0)}^{\left(1,1\right)} & = & -\frac{1}{2}\Lambda_{8}, & J_{(-\frac{1}{2},\frac{1}{2},\frac{1}{2})}^{\left(1,1\right)} & = & \frac{1}{2\sqrt{2}}\left(\Lambda_{4}+i\Lambda_{5}\right),\\
J_{(0,1,0)}^{\left(1,1\right)} & = & \frac{1}{2}\Lambda_{3}, & J_{(\frac{1}{2},\frac{1}{2},-\frac{1}{2})}^{\left(1,1\right)} & = & \frac{1}{2\sqrt{2}}\left(\Lambda_{4}-i\Lambda_{5}\right),\\
J_{(0,1,1)}^{\left(1,1\right)} & = & -\frac{1}{2\sqrt{2}}\left(\Lambda_{1}+i\Lambda_{2}\right), & J_{(-\frac{1}{2},\frac{1}{2},-\frac{1}{2})}^{\left(1,1\right)} & = & \frac{1}{2\sqrt{2}}\left(\Lambda_{6}+i\Lambda_{7}\right),\\
J_{(0,1,-1)}^{\left(1,1\right)} & = & \frac{1}{2\sqrt{2}}\left(\Lambda_{1}-i\Lambda_{2}\right), & J_{(\frac{1}{2},\frac{1}{2},\frac{1}{2})}^{\left(1,1\right)} & = & -\frac{1}{2\sqrt{2}}\left(\Lambda_{6}-i\Lambda_{7}\right).\end{array}\end{equation}

In the case of the fundamental representation $\Lambda_{k}$ matrices
reduce to the standard Gell-Mann matrices $\lambda_{k}.$

The generators $J_{(Z,I,M)}^{\left(1,1\right)}$ obey the hermitean
conjugation relation: \begin{equation}
\left(J_{\left(Z,I,M\right)}^{(1,1)}\right)^{\dagger}=\left(-1\right)^{Z+M}J_{\left(-Z,I,-M\right)}^{(1,1)}.\end{equation}
 The action of the operators $J_{(Z,I,M)}^{\left(1,1\right)}$ on
the basis states and the commutation relations are given in  \cite{jurciukonis1}.
The dimension of an arbitrary representation $\left(\lambda,\mu\right)$
is denoted by: \begin{equation}
\dim\left(\lambda,\mu\right)=\frac{1}{2}\left(\lambda+1\right)\left(\mu+1\right)\left(\lambda+\mu+2\right).\end{equation}
 The explicit expressions for the quadratic and cubic Casimir operators
are, respectively: \begin{eqnarray}
C_{2}(\lambda,\mu) & = & \frac{1}{3}\left(\lambda^{2}+\mu^{2}+\lambda\mu+3\lambda+3\mu\right),\\
C_{3}(\lambda,\mu) & = & \frac{1}{162}\left(\lambda-\mu\right)\left(\lambda+2\mu+3\right)\left(2\lambda+\mu+3\right).\nonumber \end{eqnarray}


\begin{thebibliography}{10}
\bibitem{skyrme1} T. H. R. Skyrme, A non-linear field theory, Proc.
Roy. Soc., \textbf{260}, 127 (1961)

\bibitem{manton1} N. Manton, Topological Solitons, Cam. Uni. Press,
Cambridge (2004)

\bibitem{manohar1} A. Manohar, Equivalence of the chiral soliton
and quark models in large N, Nucl. Phys. \textbf{B248}, 19 (1984)

\bibitem{prasz} M. Praszalowicz, A comment on the phenomenology of
the SU(3) Skymre model, Phys. Lett. \textbf{B158}, 264 (1985)

\bibitem{klebanov1} C. G. Callan and I. Klebanov, Bound-state approach
to strangeness in the Skyrme model, Nucl. Phys. \textbf{B262}, 365
(1985)

\bibitem{klebanov2} C. G. Callan, K. Hornbostel and I. Klebanov,
You are not entitled to access the full text of this document Baryon
masses in the bound state approach to strangeness in the skyrme model,
Phys. Lett. \textbf{B202}, 269 (1988)

\bibitem{bj} M. Bj\"ornberg, K. Dannbom and D. O. Riska, The Anharmonic
Correction in the Soliton Model of the Hyperons, Nucl. Phys. \textbf{A582},
621 (1995)

\bibitem{sco1} M. Rho, D. O. Riska and N. Scoccola, Charmed Baryons
as Soliton - D Meson Bound States, Phys. Lett. \textbf{B251}, 597
(1990)

\bibitem{sco2} M. Rho, D. O. Riska and N. Scoccola, The Energy Levels
of the Heavy Flavour Baryons in the Topological Soliton Model, Z.
Phys. \textbf{A341}, 343 (1992)

\bibitem{jurciukonis1} D. Jurciukonis, E. Norvaisas and D. O. Riska,
Canonical Quantization of SU(3) Skyrme Model in a General Representation,
Journal of Math. Phys. \textbf{46}, 072103 (2005)

\bibitem{fujii1} K. Fujjii, A. Kobushkin, K. Sato and N. Toyota,
Skyrme-model Lagrangian in quantum mechanics:SU(2) case, Phys. Rev.
\textbf{D35}, 1896 (1987)

\bibitem{fujii2} K. Fujjii, K. Sato and N. Toyota, Quantum-mechanical
aspects of SU(3) Skyrme model in collective-coordinate quantization,
Phys. Rev. \textbf{D37}, 3663 (1988)

\bibitem{acus1} A. Acus, E. Norvaisas and D.O. Riska, Stability and
Representation Dependence of the Quantum Skyrmion, Phys. Rev. \textbf{C57},
2597 (1997)
\end{thebibliography}

\end{document}